# Web of Science Core Collection's coverage expansion: The forgotten Arts & Humanities Citation Index?

Weishu LIU[1], Rong NI[2, 3]*, Guangyuan HU[2]*

1. School of Information Management and Engineering, Zhejiang University of Finance and Economics, Hangzhou 310018, Zhejiang, China.
2. School of Public Economics and Administration, Shanghai University of Finance and Economics, Shanghai 200433, Shanghai, China.
3. Centre for Science and Technology Studies (CWTS), Leiden University, Leiden 2333 BN, Netherlands.

* Corresponding author: Rong NI, School of Public Economics and Administration, Shanghai University of Finance and Economics, Shanghai 200433, Shanghai, China; Centre for Science and Technology Studies (CWTS), Leiden University, Leiden 2333 BN, Netherlands, Email: nirong007@foxmail.com; Guangyuan HU, School of Public Economics and Administration, Shanghai University of Finance and Economics, Shanghai 200433, Shanghai, China, Email: hu.guangyuan@mail.shufe.edu.cn.

**Declaration of Competing Interest**

The authors declare that they have no known competing financial interests or personal relationships that could have appeared to influence the work reported in this paper.

**CRediT authorship contribution statement**

**Weishu LIU:** Conceptualization, Investigation, Methodology, Writing – original draft, Writing – review & editing; **Rong NI:** Data curation, Formal analysis, Methodology, Writing – original draft; **Guangyuan HU:** Conceptualization, Methodology, Writing – review & editing.

# Web of Science Core Collection's coverage expansion: The forgotten Arts & Humanities Citation Index?

**Abstract** The expansion of Web of Science Core Collection (WoSCC) over the recent years has partially accounted for the "norm" of growth of research output in many bibliometric analysis studies. However, the expansion patterns of different citation indexes may be different, which may benefit some disciplines but hinder others. Utilizing Science Citation Index Expanded (SCIE), Social Sciences Citation Index (SSCI), and Arts & Humanities Citation Index (A&HCI), this study attempts to elaborate on WoSCC's coverage expansion patterns among these three databases from 2001 to 2020. Results show that different from SCIE/SSCI, both the annual publication volumes in the A&HCI database and all A&HCI journals have remained relatively stagnant in all document types considered scenario or have gained relatively slight increases in only citable items considered scenario. Although the number of A&HCI journals also has increased remarkably, the average journal publication volume of A&HCI journals has decreased gradually if all document types are considered or kept relatively stagnant when citable items only are considered. Besides, the A&HCI database has ceased the systematic index of individually selected items from SCIE/SSCI journals since 2018. The study finally discusses the possible causes and consequences of the unbalanced expansion of WoSCC's different citation indexes.

## 1. Introduction

In many bibliometric analysis studies and research evaluation practices, growth is the common "norm" regarding the volume of research output for a country or research field. Instead, a decline or stagnation in research output is often seen as abnormal. The "norm" of growth looks like a win-win-win situation for multiple stakeholders. However, the growth of research output may also be an artifact caused by the expansion of the used database itself (Harzing & Alakangas, 2016; Larsen & Von Ins, 2010; Michels & Schmoch, 2012). The prestigious Web of Science Core Collection (WoSCC), owned by Clarivate Analytics, is one of the most used bibliographic data sources in these studies and research evaluation practices (Zhu & Liu, 2020). However, for the WoSCC itself, several coverage biases exist including language, region, and discipline biases (Liu, 2017; Martín-Martín et al., 2018, 2021; Mongeon & Paul-Hus, 2016; Singh et al., 2021; Vera-Baceta et al., 2019), which are critical limitations of WoSCC. To reduce these biases, WoSCC has rapidly expanded its coverage of scholarly contents (Birkle et al., 2020; Testa, 2011).

Along with the growth of WoSCC, Vanderstraeten & Vandermoere (2021) probed the inequality of discipline expansion among social sciences via using the Social Sciences Citation Index (SSCI). However, different citation indexes of WoSCC may also show different expansion patterns. Uneven expansion among different citation indexes may benefit some disciplines, however, it may also hinder the development of others. Therefore, special attention should be paid to the data analysis, result interpretation and application when using WoSCC as the data source. In this study, we took all disciplines from natural science, and social sciences to arts & humanities into consideration and focused on the different expansion patterns from the macro perspective of citation index. One of the key points is to identify individually selected papers/journals in each citation index to dig into the causes behind the different expansion patterns of the three flagship journal citation indexes.

As WoSCC claims, it covers journals from all disciplines and worldwide regions and only includes journals that exhibit high-level editorial rigor and best practice[1]. Studies have shown an exponential growth in papers indexed by WoSCC from 1990 to 2019 (Hu, Leydesdorff & Rousseau, 2020). The growth of WoSCC-indexed papers is associated with both the growth of science and additional coverage of used databases (Michels & Schmoch, 2012). Meanwhile, extant literature also points out the biases existing in the development of WoSCC. For example, previous studies have found the unbalanced coverage of WoSCC among natural science, social sciences, and arts & humanities, which is conducive to natural sciences while detrimental to social sciences and the humanities (Aksnes & Sivertsen, 2019; Hicks & Wang, 2014; Leydesdorff, Hammarfelt & Salah, 2011; Mongeon & Paul-Hus, 2016). Moreover, although WoSCC began to include books and chapters (Book Citation Index since 2005) (Liu, 2019; Torres-Salinas et al., 2013, 2014), it did not fully overcome the discipline bias. In addition, English language journals have an overwhelming advantage regarding the coverage and citation impact over non-English journals in WoSCC, resulting in the underestimation of contribution of non-English research output to academic research (Archambault et al., 2006; Liu, 2017; Liu et al., 2018). Similar actions included the large-scale coverage of top-tier regional journals during 2007-2010 (Testa, 2011) and the introduction of the Emerging Sources Citation Index in 2015 (Huang et al., 2017; De Filippo & Gorraiz, 2020).

All these efforts taken by WoSCC are worthy of recognition. However, the expansion patterns of different citation indexes are still not clear to the scientific community. The unbalanced expansion between different citation indexes may benefit some disciplines but hinder others. This paper contributes to revealing inequalities in the expansion of different citation indexes in WoSCC. In contrast to the previous studies, this study employs a method to reclassify journals by matching and comparison, thus distinguishing between one citation index's own fully indexed journals and individually selected journals. The findings not only reveal the uncommon characteristics of the growth of papers and journals in A&HCI database, but also probe the possible reasons behind the dynamics through a comparative analysis. This empirical analysis also provides practical insights for research evaluation.

Specifically, we address the following research questions in this paper:

1. What are the expansion models of papers and journals on WoSCC's three main journal citation indexes when including and excluding individually selected items?

2. What's the impact of annual journal publication volumes on database expansion?

3. What's the potential impact of OA journal on journal volume/database expansion?

## 2. Data and methodology

### 2.1 Data

The data source of this study is three classical journal citation indexes of WoSCC (i.e., Science Citation Index Expanded (SCIE), Social Sciences Citation Index (SSCI), and Arts & Humanities Citation Index (A&HCI)). Data for the period of 2001 to 2020 were collected. The data were accessed on October 15th, 2021. Since the publication volumes of different citation indexes varied greatly, data of the year 2001 were chosen as the benchmark to calculate the growth rates of the number of papers or journals in each citation index.

---

[1] https://mjl.clarivate.com/home

It is noteworthy that as online-first papers have been indexed by WoSCC since December 2017, two versions of publication year values exist for some ever-early access records (Liu, 2021). In this study, we adopted the "publication years" on the result analysis page of WoSCC to identify the number of papers/journals of each year.

**2.2 Indicators**

***SCIE (or SSCI, or A&HCI) journals:*** According to the Master Journal List provided by Clarivate[2], each journal indexed by WoSCC belongs to one or more journal citation indexes. An SCIE (or SSCI, or A&HCI) journal refers to a journal fully indexed by SCIE (or SSCI, or A&HCI) database in one specific year, and generally, all papers published by this journal in this year are fully indexed in SCIE (or SSCI, or A&HCI) database.

***SCIE (or SSCI, or A&HCI) journal papers:*** They refer to papers published by SCIE (or SSCI, or A&HCI) journals.

***Journals retrieved from SCIE (or SSCI, or A&HCI) database:*** When searching papers from WoSCC, journals retrieved from SCIE/SSCI/A&HCI database are called *Journals retrieved from SCIE/SSCI/A&HCI database* respectively. It is worth noting that *Journals retrieved from SSCI database* are not certainly *SSCI journals* since the SSCI database also individually selects some social science-related items from outstanding natural science journals. Similarly, *journals retrieved from A&HCI database* are also not necessarily *A&HCI journals*[3].

***Papers retrieved from SCIE (or SSCI, or A&HCI) database:*** As mentioned above, they refer to papers retrieved from SCIE (or SSCI, or A&HCI) database.

***Individually selected papers (or journals)[4]:*** As described in the help file of WoSCC, "SSCI also indexes individually selected, relevant items of the world's leading scientific and technical journals"[5] and "A&HCI also indexes individually selected, relevant items from major science and social science journals". Accordingly, we defined these individually selected papers in each database as *individually selected papers* and corresponding journals publishing these papers as *individually selected journals* in each database. Table 1 describes three journal citation indexes' coverage regarding these two types of journals for one year.

***Journal publication volume:*** *Journal publication volume (JPV)* represents the annual number of papers in a journal. In this study, we used the number of papers published in a journal per year exported from WoSCC as the value of JPV[6].

**Table 1.** Three citation indexes' journal coverage for one year

---

[2] The Master Journal List is a tool for researchers to find and identify the right journal on the Web of Science platform. The list of journals is updated on an at least monthly basis. See https://mjl.clarivate.com/home.

[3] http://webofscience.help.clarivate.com/en-us/Content/wos-core-collection/wos-core-collection.htm.

[4] "Individually selected journal" is named by the authors themselves. This study chooses to use this terminology for the convenience of explanation .

[5] Data show that SSCI database also individually selects a limited number of items from A&HCI journals for some years. For example, the A&HCI journal *Style* published 44 papers in 2006 and two of them were individually selected into the SSCI database.

[6] Although the Journal Master List contains citable item indicator for SCIE and SSCI journals from 1997 to 2020 (Vanderstraeten & Vandermoere, 2021), it does not include corresponding data for A&HCI journals. In order to unify the calculation method, this study does not use the citable item indicator directly but uses journal publication volume exported from WoSCC. See: http://help.incites.clarivate.com/incitesLiveJCR/9607-TRS/version/9.

| Journal Citation Index | Journal coverage |
|---|---|
| SCIE database | Journals only fully indexed by SCIE |
| | Journals fully indexed by SCIE, and also fully indexed by SSCI or A&HCI |
| SSCI database | Journals only fully indexed by SSCI |
| | Journals fully indexed by SSCI, and also fully indexed by SCIE or A&HCI |
| | **Individually selected journals from SCIE or A&HCI journals** |
| A&HCI database | Journals only fully indexed by A&HCI |
| | Journals fully indexed by A&HCI, and also fully indexed by SCIE or SSCI |
| | **Individually selected journals from SCIE or SSCI journals** |

*Notes:* a. We also examined whether the SCIE database included individually selected papers from SSCI journals. The results showed that only a minimal number of papers in the SCIE database were individually selected from SSCI journals. b. After checking, there were no individually selected A&HCI journals in the SCIE database. Therefore, the following corresponding analyses would neglect these individually selected items in the SCIE database.

## 2.3 Matching methods

To uncover the expansion patterns of different citation indexes of WoSCC, the indicators mentioned above should be identified clearly. For the analyses of "SSCI journal papers" and "A&HCI journal papers", individually selected journals in SSCI/A&HCI database should be distinguished and eliminated. Based on the journal coverage description in Table 1, Fig. 1 illustrates the matching methods and recognition rules of individually selected journals in the SSCI/A&HCI database used in this investigation.

Firstly, the yearly "Publication titles" (PT) and "Record count" (RC) of journals retrieved from SCIE, SSCI, and A&HCI databases were exported respectively from the search results pages of WoSCC. Corresponding variables for the year y were named as $PT\_SCIE_y$, $RC\_SCIE_y$, $PT\_SSCI_y$, $RC\_SSCI_y$, $PT\_A\&HCI_y$, $RC\_A\&HCI_y$ respectively. To identify the individually selected items in the SSCI database, $PT\_SCIE_y$ and $PT\_A\&HCI_y$ were matched with $PT\_SSCI_y$. Thus, we obtained the numbers of papers published by the same journal but indexed in the SSCI, SCIE, and A&HCI databases for the year y respectively. By comparing the values of $RC\_SSCI_y$ with $RC\_SCIE_y$ and $RC\_A\&HCI_y$, individually selected journals in the SSCI database from SCIE and A&HCI journals can be identified.

Fig. 1(a) demonstrates the identification of individually selected journals in the SSCI database. For an individually selected journal in the SSCI database, the SSCI database generally does not index all papers published in this journal. Therefore, if the $RC\_SSCI_y$ of a journal is smaller than its $RC\_SCIE_y$, we determine that it is not an SSCI journal for the year y[7]. If the $RC\_SSCI_y$ of a journal is larger than or equal to its $RC\_SCIE_y$, we subsequently compare the values of $RC\_SSCI_y$ and $RC\_A\&HCI_y$. For the case of "$RC\_SSCI_y < RC\_A\&HCI_y$", we believe that the corresponding journal should be individually selected rather than an SSCI journal; otherwise, it is an SSCI journal. A similar identification of individually selected journals in the A&HCI database is illustrated in Fig. 1(b). Random inspections and corrections were conducted to mitigate possible misclassification. For example, for journals whose publication records satisfy "$RC\_SSCI_y = RC\_SCIE_y$ or $RC\_SSCI_y = RC\_A\&HCI_y$" and "$RC\_SSCI_y \leqq 5$", we searched corresponding journals in the Journal Citation Reports for further adjustments and corrections. Similar verification work was conducted after distinguishing individually selected journals in the A&HCI database.

---

[7] In this paper, we only need to identify whether journals retrieved from SSCI database are SSCI journals or individually selected journals from SCIE/A&HCI journals, therefore, we do not further distinguish whether these SSCI journals are also SCIE or A&HCI journals.

We further took the prestigious and representative periodical Journal of Informetrics (JoI) as an example to verify the above identification methods. Journal Citation Reports show that JoI has been an SSCI journal since 2008, and has been both SCIE and SSCI journal since 2016. We also compared the annual publication data exported from WoSCC year by year using the identification methods demonstrated in Fig. 1. The results showed that JoI has been recorded in the SSCI database since 2007 and has belonged to SSCI journals since then. Since 2016, all papers in JoI have been fully indexed in both SSCI and SCIE databases, indicating JoI has been both an SCIE and SSCI journal since then. After considering the retrospective data inclusion of newly indexed journals, this conclusion is consistent with the data from the Journal Citation Reports.

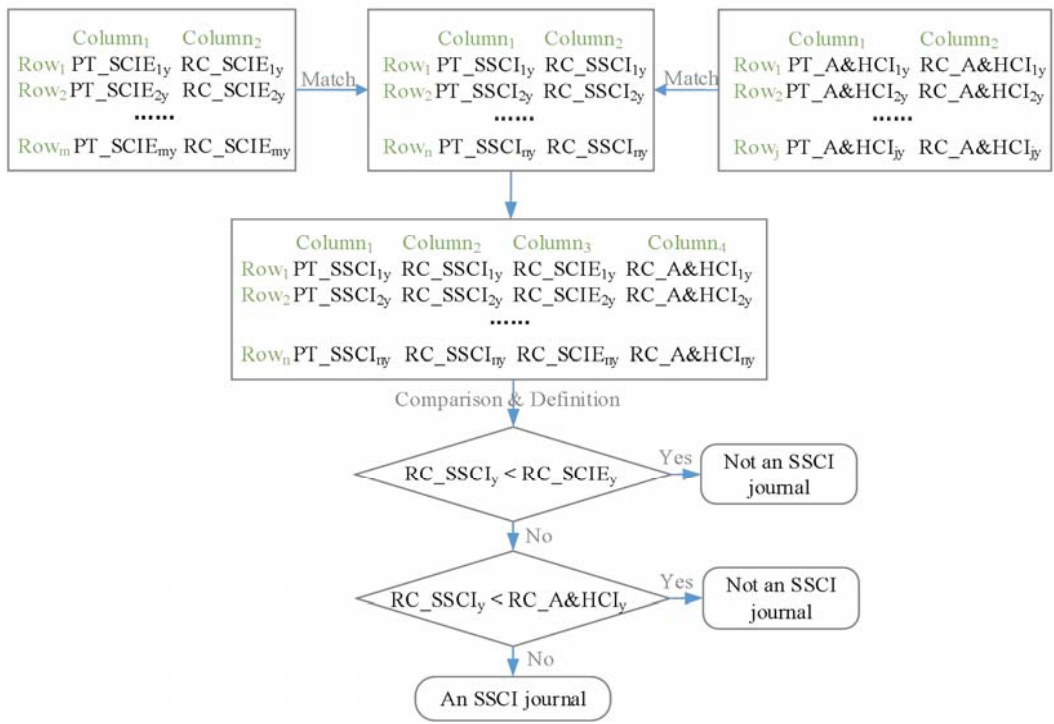

(a) Journal reclassification in SSCI database for the year y

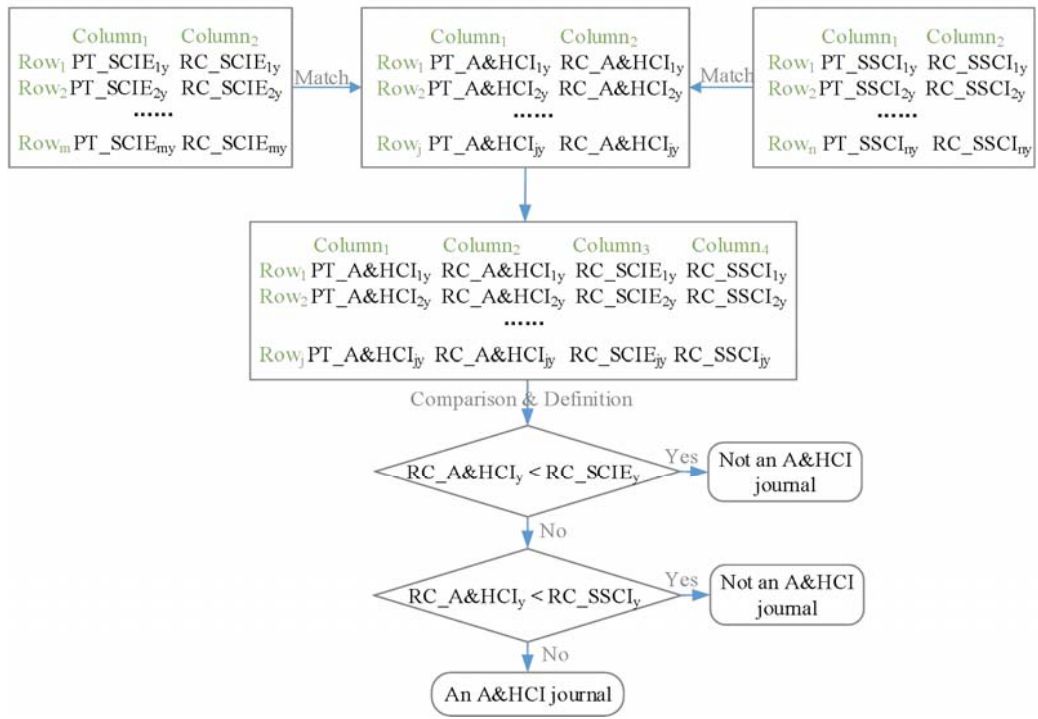

(b) Journal reclassification in A&HCI database for the year y

**Fig. 1.** Matching and reclassification methods

## 3. Analyses

**3.1 Web of Science Core Collection's coverage expansion: an overview**

*Publication volume growth*

Fig. 2 depicts the publication volume growth of three flagship journal citation indexes of WoSCC during the study period respectively. For all document types considered scenario in Fig. 2(a), it can be easily found that both the numbers of papers retrieved from SCIE and SSCI databases showed significant growth. In 2020, the number of papers retrieved from the SCIE database was 2.45 times that of the year 2001. The number of SSCI papers has increased rapidly since 2008 and finally tripled in 2020. Contrarily, the number of papers retrieved from the A&HCI database has hardly changed over the past 20 years, and remained even in a stage of declining from 2002 to 2004.

Since the A&HCI database contains a large proportion of non-citable items such as book reviews (Liu et al., 2017), we further restricted the document types only to citable items (i.e., articles and review articles) which are considered as "the substantive articles that contribute to the body of scholarship"[8]. As Fig. 2(b) shows, the numbers of citable items in all three databases also followed growth trends. Similar to the trends demonstrated in Fig. 2(a), the number of articles and review articles retrieved from the SCIE/SSCI database in 2020 was 2.64/4.45 times that of the year 2001 respectively. Compared with all document types considered in the A&HCI database, the number of citable items in the A&HCI database has gained a slight increase over the 20 years from roughly 31,000 in 2001 to 54,000 in 2020 with an increase of 71%.

---

[8] https://help.incites.clarivate.com/incitesLiveJCR/9607-TRS

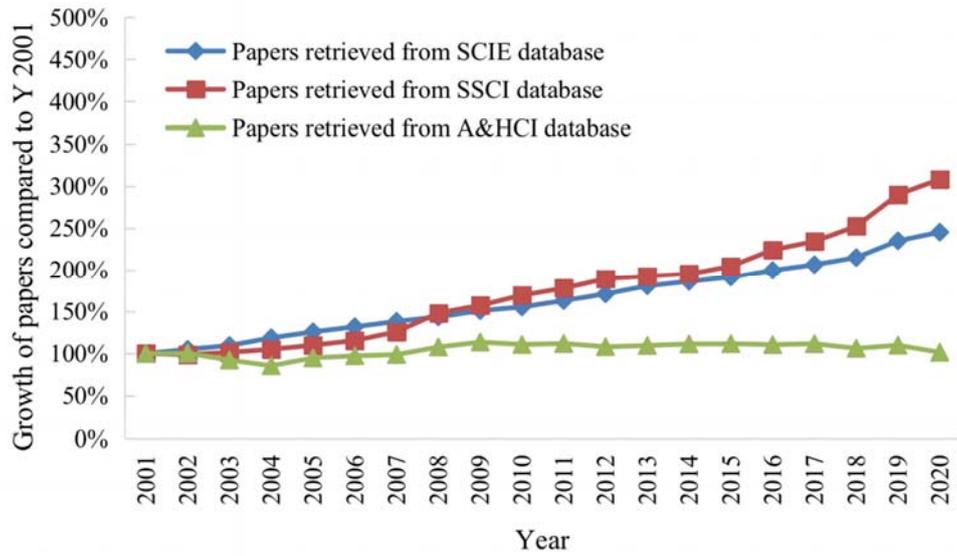

(a) All document types

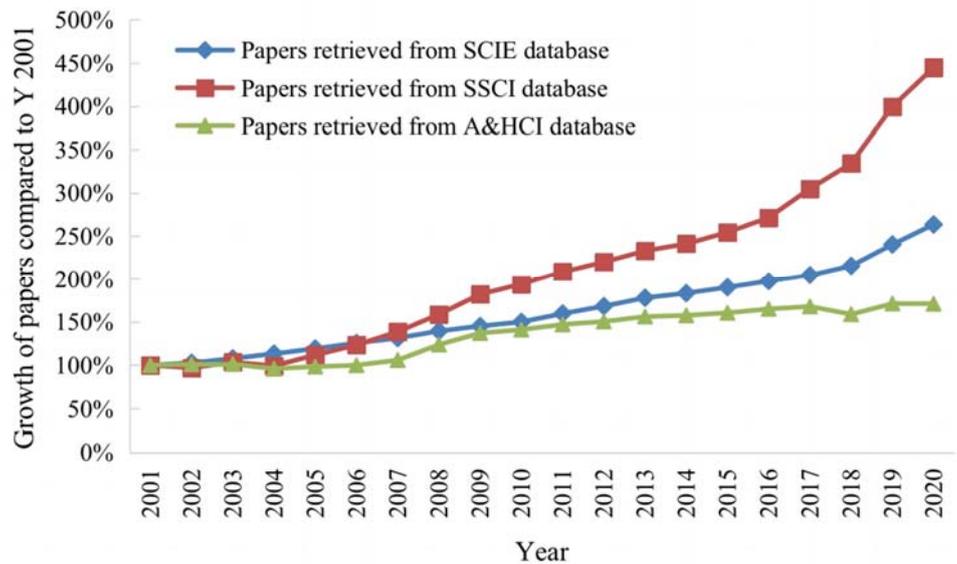

(b) Articles and review articles

**Fig. 2.** Dynamics of papers in WoSCC's three citation indexes

*Notes:* Growth of papers= Number of papers in a specific year/Number of papers in the year 2001. The data were accessed from Fudan University Library on 15 October 2021.

*Journal growth*

Fig. 3 shows that the numbers of publication titles retrieved from three databases display different growth patterns[9]. The number of publication titles retrieved from the SCIE database has grown relatively slowly from about 6500 in 2001 to 9700 in 2020, with an increase of about 48%[10]. Comparatively, the number of publication titles in the SSCI database has expanded rapidly from about 3500 in 2001 to 8100 in 2020, with an increase of 128%. Nevertheless, the number of publication titles in the A&HCI database appeared to be irregular, roughly showing a fluctuating trend. From 2001 to 2009, the number of publication titles in the A&HCI database showed an overall increasing trend; however, after 2010 (especially after 2017), it showed an overall decreasing trend, which may be related to the distortion of individually selected journals in the A&HCI database.

---

[9] There is a small share of records with both journal titles and book series titles in the WoSCC, which results in a slight over-counting of the number of papers and journals. For example, when searching for the journal *Annual Review of Virology*, both *Annual Review of Virology* and *Annual Review of Virology VOL\** appear on the search results page.

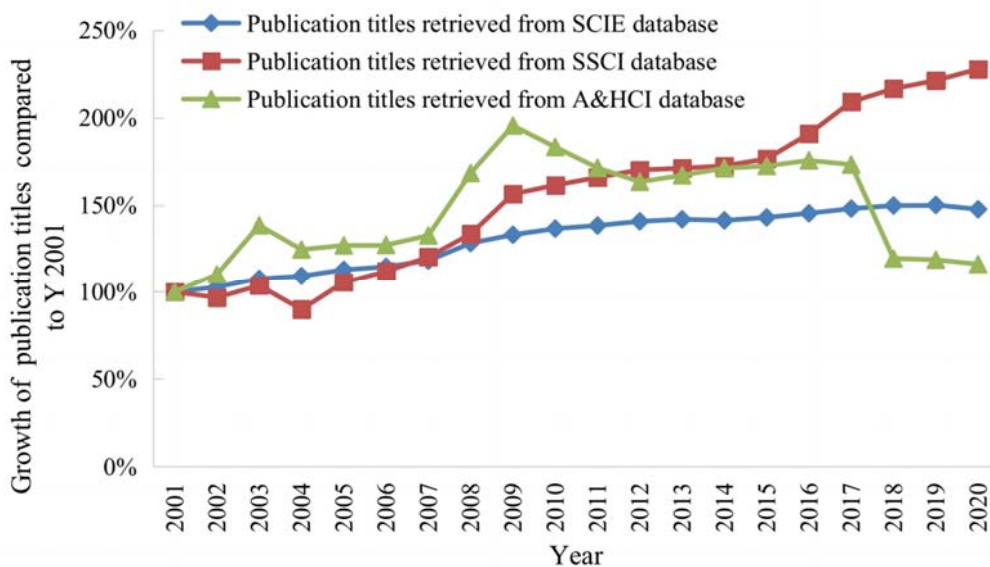

**Fig. 3.** Growth trends of the numbers of publication titles retrieved from three databases

*Notes*: Growth of publication titles = Number of publication titles in a specific year/Number of publication titles in the year 2001.

### 3.2 Individually selected items

*Dynamics of individually selected papers/journals*

As mentioned earlier, journals retrieved from the A&HCI database are not equivalent to A&HCI journals. These journals may also be individually selected from SSCI or SCIE journals. According to the definition rules and matching methods of this study, individually selected papers and corresponding journals in SSCI and A&HCI databases were identified respectively. The distribution of individually selected items in SSCI and A&HCI databases is shown in Table 2 (all document types considered).

It is found that increasing numbers of SCIE/A&HCI journals and corresponding papers were individually selected into the SSCI database with only a few exceptional years. For each of the 20 years, about half of the journals retrieved from the SSCI database were individually selected. The proportion of individually selected papers among the total papers retrieved from the SSCI database has also increased gradually. In 2001, the share was only 9.6%, increasing to 23.9% by 2020. One of the reasons for the rapid increase of papers in the SSCI database could be the growing number of individually selected papers from SCIE/A&HCI journals.

---

[10] A slight decrease of the number of publication titles retrieved from SCIE database can be witnessed in Fig. 3 for the year 2020. It is mainly due to the delayed indexation of some book series titles and retrospective data for some newly indexed journals.

Table 2 shows that the A&HCI database has individually selected papers from SCIE or SSCI journals systematically until 2017. From 2001 to 2017, about 1/3 of journals in the A&HCI database were individually selected. However, less than 4% of papers indexed in the A&HCI database were from these individually selected journals. Interestingly, for the year 2018, only 135 papers in the A&HCI database were individually selected from eight SCIE or SSCI journals. Individually selected journals disappeared completely from the A&HCI database in 2020. The disappearance of individually selected papers and journals in the A&HCI database indicates a sudden shift of inclusion policy of individually selected items in the A&HCI database.

**Table 2.** Individually selected journals in SSCI and A&HCI databases (all document types)

| | SSCI database | | | | A&HCI database | | | |
| | Individually selected journal | | Individually selected papers | | Individually selected journal | | Individually selected papers | |
| Year | # | % | # | % | # | % | # | % |
| --- | --- | --- | --- | --- | --- | --- | --- | --- |
| 2001 | 1747 | 48.8% | 13834 | 9.6% | 420 | 27.4% | 809 | 0.7% |
| 2002 | 1624 | 46.9% | 11355 | 8.0% | 573 | 33.9% | 1867 | 1.7% |
| 2003 | 1847 | 49.7% | 14694 | 10.0% | 1002 | 47.2% | 3221 | 3.1% |
| 2004 | 1351 | 41.9% | 8672 | 5.7% | 781 | 40.8% | 2106 | 2.2% |
| 2005 | 1807 | 47.8% | 15596 | 9.8% | 802 | 41.2% | 2330 | 2.2% |
| 2006 | 1956 | 48.7% | 17378 | 10.4% | 791 | 40.5% | 1858 | 1.7% |
| 2007 | 1979 | 45.9% | 18478 | 10.2% | 783 | 38.5% | 2050 | 1.9% |
| 2008 | 1983 | 41.5% | 17416 | 8.1% | 1046 | 40.5% | 3029 | 2.5% |
| 2009 | 2561 | 45.7% | 28345 | 12.5% | 1363 | 45.4% | 4526 | 3.6% |
| 2010 | 2639 | 45.6% | 31390 | 12.8% | 1116 | 39.7% | 4175 | 3.4% |
| 2011 | 2708 | 45.6% | 33516 | 13.1% | 904 | 34.4% | 3652 | 2.9% |
| 2012 | 2821 | 46.3% | 37135 | 13.6% | 757 | 30.2% | 3289 | 2.7% |
| 2013 | 2848 | 46.5% | 39887 | 14.4% | 803 | 31.3% | 3597 | 2.9% |
| 2014 | 2889 | 46.7% | 40349 | 14.3% | 853 | 32.5% | 3865 | 3.1% |
| 2015 | 2944 | 46.6% | 42453 | 14.4% | 843 | 31.9% | 4113 | 3.3% |
| 2016 | 3396 | 49.7% | 52645 | 16.3% | 884 | 32.8% | 4455 | 3.6% |
| 2017 | 3954 | 52.8% | 69862 | 20.7% | 821 | 30.9% | 3671 | 2.9% |
| 2018 | 4189 | 54.0% | 80743 | 22.2% | 8 | 0.4% | 135 | 0.1% |
| 2019 | 4353 | 54.9% | 93531 | 22.4% | 3 | 0.2% | 84 | 0.1% |
| 2020 | 4623 | 56.6% | 106266 | 23.9% | 0 | 0.0% | 0 | 0.0% |

*Notes*: # means the number of individually selected journals/papers. % means the share of individually selected journals/papers among SSCI/A&HCI database.

*Dynamics of papers excluding individually selected ones*

To further uncover the patterns behind the expansion of the three databases, the analyses were conducted after excluding individually selected items. Fig. 4(a) presents the dynamics of papers of all document types after excluding individually selected ones. As also depicts in Fig. 2(a), SCIE journal papers exhibited consistent growth. In comparison, SSCI journal papers' growth trend decreased slightly after excluding those individually selected papers (an increase of 159% in Fig. 4(a) vs 208% in Fig. 2(a)). The change in A&HCI journal papers was also insignificant and showed almost stagnant growth.

As Fig. 4(b) displays, document types were limited to citable items. After excluding individually selected papers, the numbers of citable items retrieved from SCIE, SSCI, and A&HCI journals in 2020 were 2.64 times, 3.86 times, and 1.75 times that of the year 2001 respectively. The expansion patterns of citable items published in A&HCI journals are different from the patterns demonstrated in all document types considered scenario. Since 2008, the number of citable items from A&HCI journals has shown an upward trend and eventually with an increase of 75% in 2020.

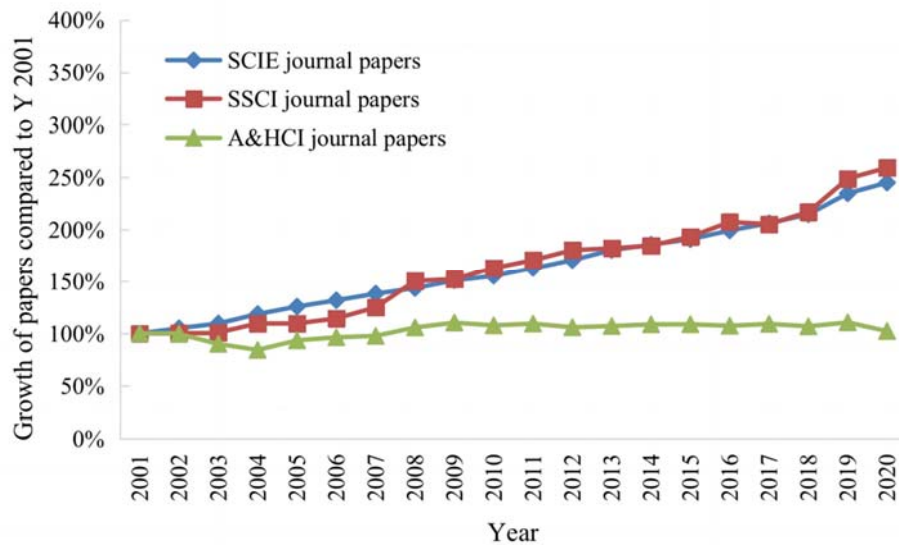

(a) All document types

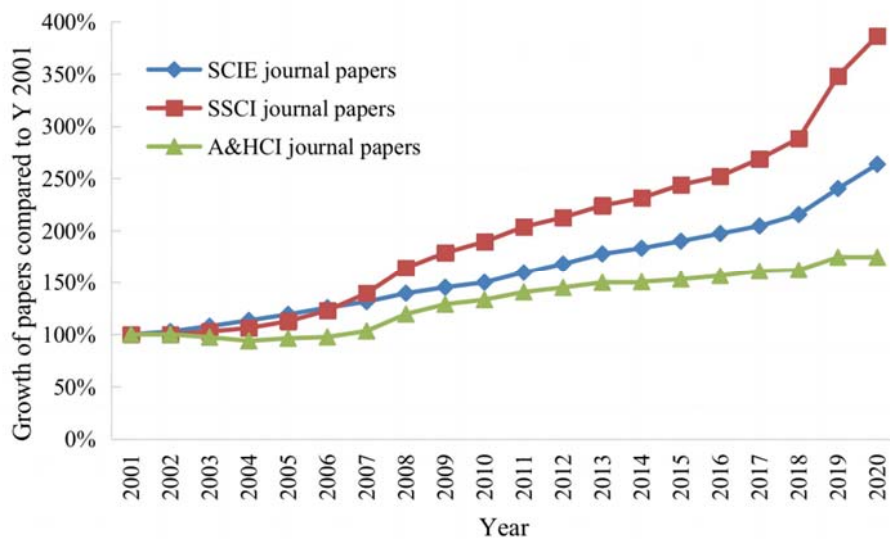

(b) Articles and review articles

**Fig. 4.** Growth trends of SCIE/SSCI/A&HCI journal papers

**Notes:** The indicators are calculated after excluding individually selected items. Growth of papers = Number of papers in a specific year/Number of papers in the year 2001.

*Dynamics of journals excluding individually selected ones*

After excluding individually selected journals according to the methods in Fig. 1, Fig. 5

demonstrates the dynamics of the numbers of SCIE, SSCI, and A&HCI journals from 2001 to 2020. Compared with the year 2001, all the numbers of SCIE, SSCI, and A&HCI journals increased significantly. The number of SCIE journals has increased relatively steadily with an increase of 48%[11]. Similarly, the number of SSCI journals in 2020 nearly doubled (an increase of 93%) compared to the year 2001. About an increase of 60% could also be witnessed for the number of A&HCI journals. A significant increase of SSCI and A&HCI journals can be witnessed during 2007-2010 which echoes the regional journal expansion initiative of the Web of Science Core Collection (Testa, 2011). However, the number of A&HCI journals remained stable over the next ten years.

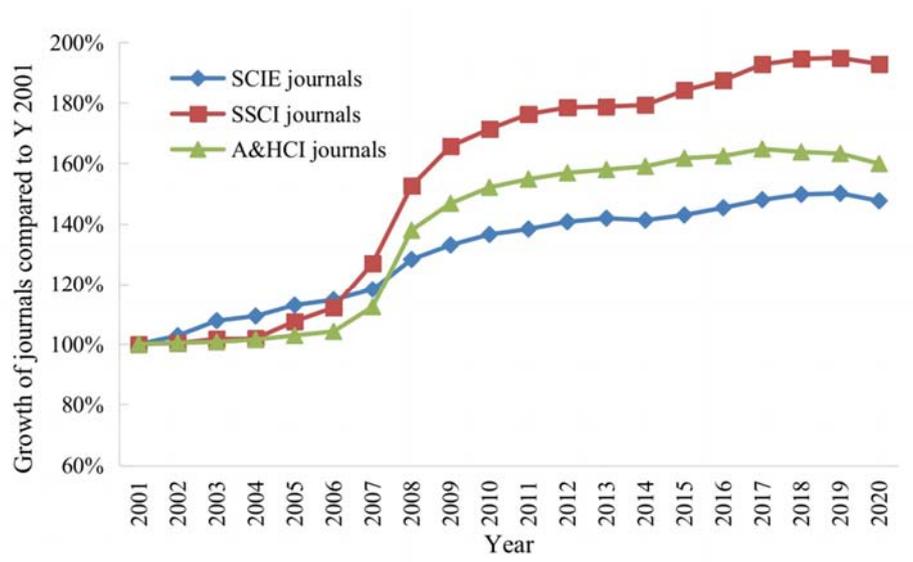

**Fig. 5.** Growth trends of SCIE/SSCI/A&HCI journals

*Notes*: The indicators are calculated after excluding individually selected items. Growth of journals = Number of journals in a specific year/Number of journals in the year 2001.

**3.3 Journal publication volumes**

*Dynamics of journal publication volumes*

To further uncover the patterns behind the changes in publication volumes, the indicator journal publication volume (JPV) is introduced in this section. Fig. 6 depicts the dynamics of average journal publication volumes (JPVs) of SCIE/SSCI/A&HCI journals from 2001 to 2020.

Fig. 6(a) demonstrates the dynamics of average JPVs of SCIE/SSCI/A&HCI journals for all document types considered scenario. The average JPVs of SCIE journals were always much higher than those of SSCI/A&HCI journals. The average JPVs of SCIE journals rose gradually from about 150 in 2001 to 250 in 2020. Comparatively, the average JPVs of SSCI journals increased a bit slower from about 70 in 2001 to about 100 in 2020. Meanwhile, there was a short period of decline after 2008, which is consistent with the description in Vanderstraeten & Vandermoere (2021). It may be partly due to the rapid increase of newly indexed regional journals, among which many small-volume journals were added (a similar trend can also be found for SCIE journals). Surprisingly, Fig. 6(a) also shows that the average JPVs of A&HCI journals were slowly decreasing from about 100

---

[11] As mentioned before, data of the last year (year 2020) will be underestimated due to delayed indexation. Non-standardized publication titles (journal titles and book series titles) for a limited share of papers have not been unified.

in 2001 to about 65 in 2020.

For the citable items considered scenario, Fig. 6(b) shows the dynamics of the average JPVs of SCIE, SSCI, and A&HCI journals. A bit different from the patterns depicted in Fig. 6(a), although the average JPVs of SCIE and SSCI journals still followed gradual upward trends, the average JPVs of A&HCI journals fluctuated in a small range. The average JPVs of SCIE journals were much larger than those of SSCI journals. A bit different from Fig. 6(a), the average JPVs of SSCI journals were also a bit larger than those of A&HCI journals as demonstrated in Fig. 6(b). This may be partly due to a large share of non-citable items in A&HCI journals (Liu et al., 2017).

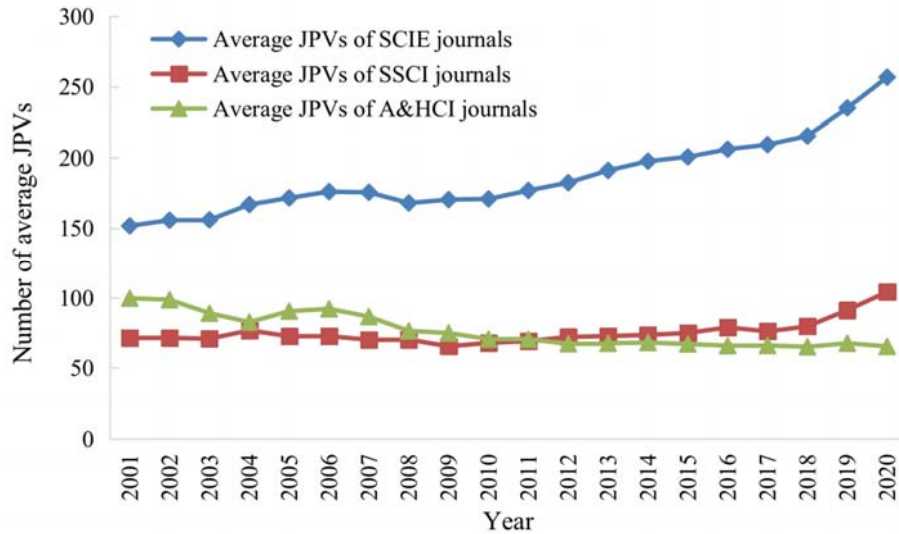

(a) All document types

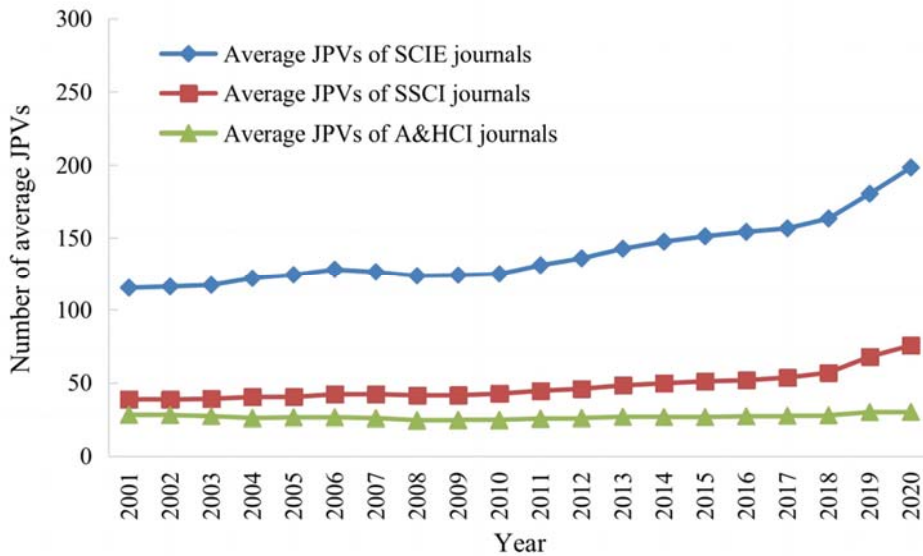

(b) Articles and review articles

**Fig. 6.** Dynamics of average journal publication volumes of SCIE/SSCI/A&HCI journals

***Growth rates of journal publication volumes***

For a better comparison of the growth dynamics of the average JPVs among SCIE, SSCI, and A&HCI journals, the growth rates of average JPVs compared to the year 2001 were also calculated. For all document types considered scenario as depicted in Fig. 7(a), it is easy to find that the average JPVs of SCIE and SSCI journals fluctuated during the first ten years but followed an upward trend since then, especially for the last three years. The rapid expansion of average JPVs may be due to the rise of online open access journals which has no strict page limit. However, although the number of A&HCI journals also increased slightly as demonstrated in Fig. 5, the average JPVs of A&HCI journals have followed a downward trend. More specifically, the average JPV of A&HCI journals in 2020 was only 2/3 of that in 2001. It is one important factor that has caused the stagnation of publication volume in the A&HCI database in the past two decades.

However, the growth patterns of average JPVs of SCIE/SSCI/A&HCI journals changed when only considering the citable items. The average JPVs of SCIE/SSCI journals followed a similar growth pattern. Significant growths can be witnessed for SCIE/SSCI journals for the last three years. However, the average JPVs of A&HCI journals first dropped for three years and then fluctuated for about ten years.

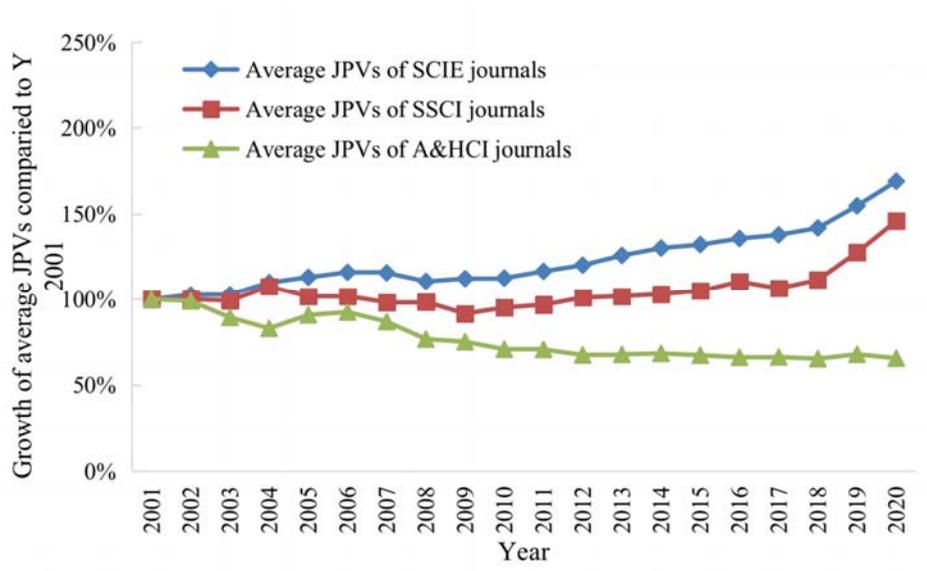

(a) All document types

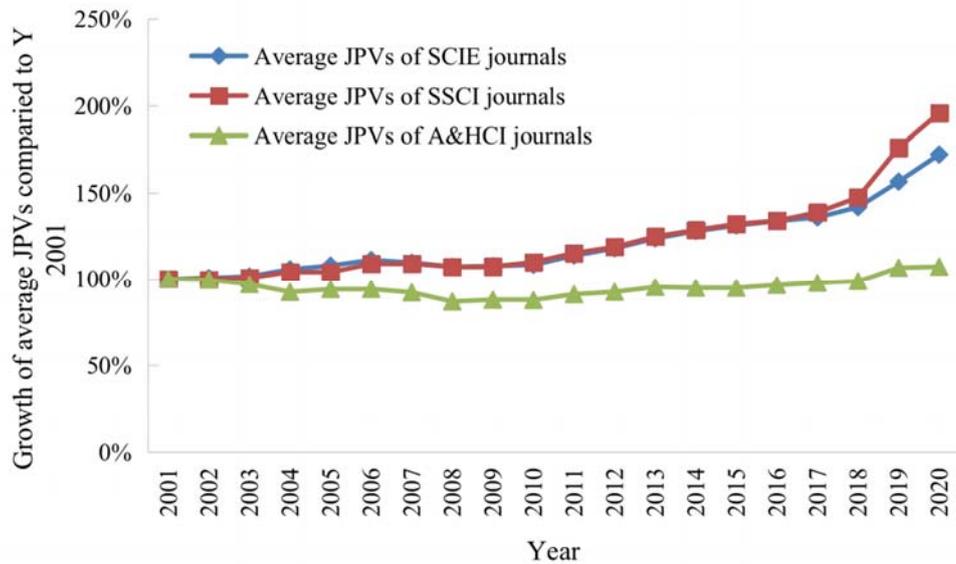

(b) Articles and review articles

**Fig. 7.** Growth trends of average JPVs

*Notes*: Individually selected items were excluded. Growth of average JPVs = Number of average JPVs in a specific year/Number of average JPVs in the year 2001.

**3.4 Open access journals**

*Proportion of gold open access journals*

As an alternative publishing model, Open Access (OA) gets remarkable advocation in scholarly communication activities in recent years (Björk, 2017; Liu & Li, 2018). Gold OA is considered as a disruptive innovation due to the characteristic that papers are available to the public without the restriction of subscription (Björk et al., 2010; Lewis, 2012). Compared with the limited annual publication volume of traditional printed journals, online open access journals have no annual publication volume restriction theoretically. Therefore, the open access mega-journals have flourished in recent years and attracted the attention of academia (Liu, 2020; Spezi et al., 2017). The growth of average JPVs of many journals in recent years may be influenced by open access journals, especially mega-ones.

We refined the "Open Access" filter on the left side of the WoSCC search results page to "Gold OA" and exported the search results by publication titles[12]. Using the same match method as demonstrated in Fig. 1, individually selected papers and corresponding journals in SSCI and A&HCI databases were excluded in corresponding analyses. Fig. 8 depicts the proportions of gold OA journals among SCIE, SSCI, and A&HCI journals respectively. Obviously, the proportion of gold OA journals among all SCIE journals rose rapidly, reaching about 1/5 in 2020. In contrast, both the proportions of gold OA journals among all SSCI and A&HCI journals grew much slower, accounting for 9% and 7% respectively in 2020.

---

[12] WoSCC includes the following six OA types: gold, gold hybrid, free to read, green publisher, green accept, and green submitted. As stated in the help file, "All articles in these journals must have a license in accordance with the Budapest Open Access Initiative to be called Gold". See https://webofscience.help.clarivate.com/en-us/Content/open-access.html.

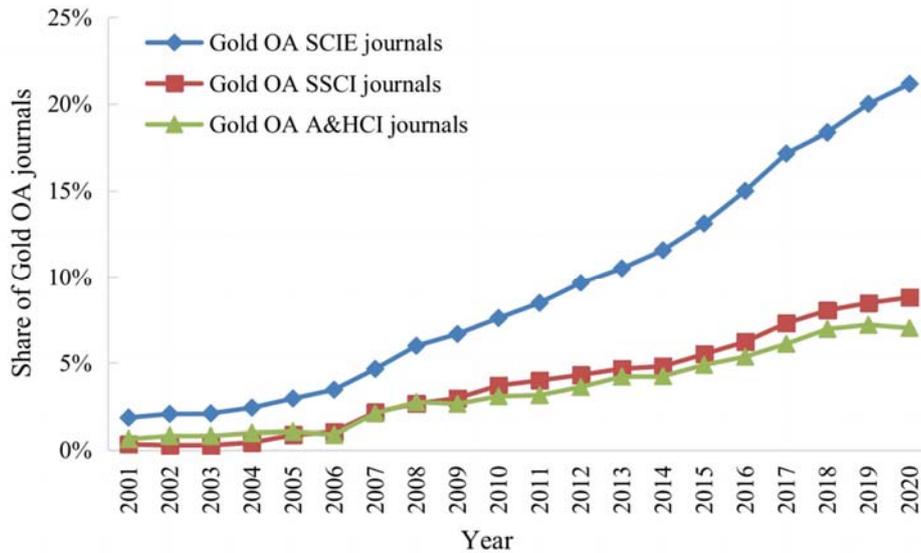

**Fig. 8.** Share of Gold OA journals among SCIE/SSCI/A&HCI journals

*Average journal publication volume of gold OA journals*

Fig. 9(a) plots the dynamics of the average JPVs of gold OA journals and all journals for SCIE/SSCI/A&HCI journals from 2001 to 2020 respectively for all document types considered scenario. It is straightforward to find that the average JPVs of gold OA SCIE and SSCI journals increased steadily yearly, but those of A&HCI journals were relatively unstable. Previous studies showed that a small number of gold OA SCIE and SSCI journals publish a large number of records per year (Liu, 2020). For example, in 2020, the SCIE journal *Scientific Reports* published about 22400 papers, the SCIE journal *PLoS One* published about 16600 papers, and SCIE/SSCI journal *Sustainability* published about 10600 papers. Nevertheless, for A&HCI journals, their average JPVs followed a downward trend (The gold OA A&HCI journal with the largest number of publications in 2020 is *Religions*, with about 650 papers). The average JPVs for gold OA A&HCI journals have been consistently much smaller than the average JPVs of all A&HCI journals. This is one of the reasons why the annual publication volume of the A&HCI database has not increased in nearly 20 years.

For the citable item considered scenario as shown in Fig. 9(b), the differences in average JPVs between all A&HCI journals and gold OA A&HCI journals narrowed significantly. It implies that articles and review articles are the primary document types in gold OA A&HCI journals, however, other A&HCI journals contain a large share of non-citable items. In recent years, the annual publication volumes of OA journals are all larger than those of ordinary journals, which is very evident for SSCI journals. Fig. 9(b) also illustrates that, unlike the gold OA SCIE and SSCI journals, gold OA A&HCI journals did not have the advantage of large publication volume, which may also be one of the causes for the stagnation of A&HCI database.

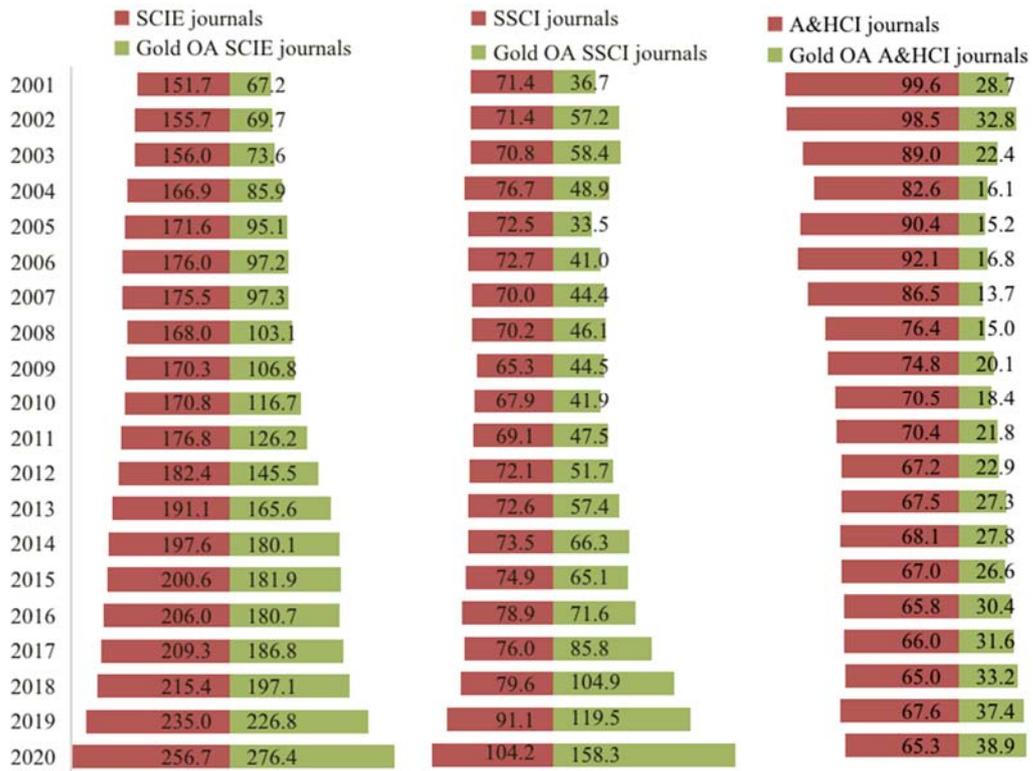

(a) All document types

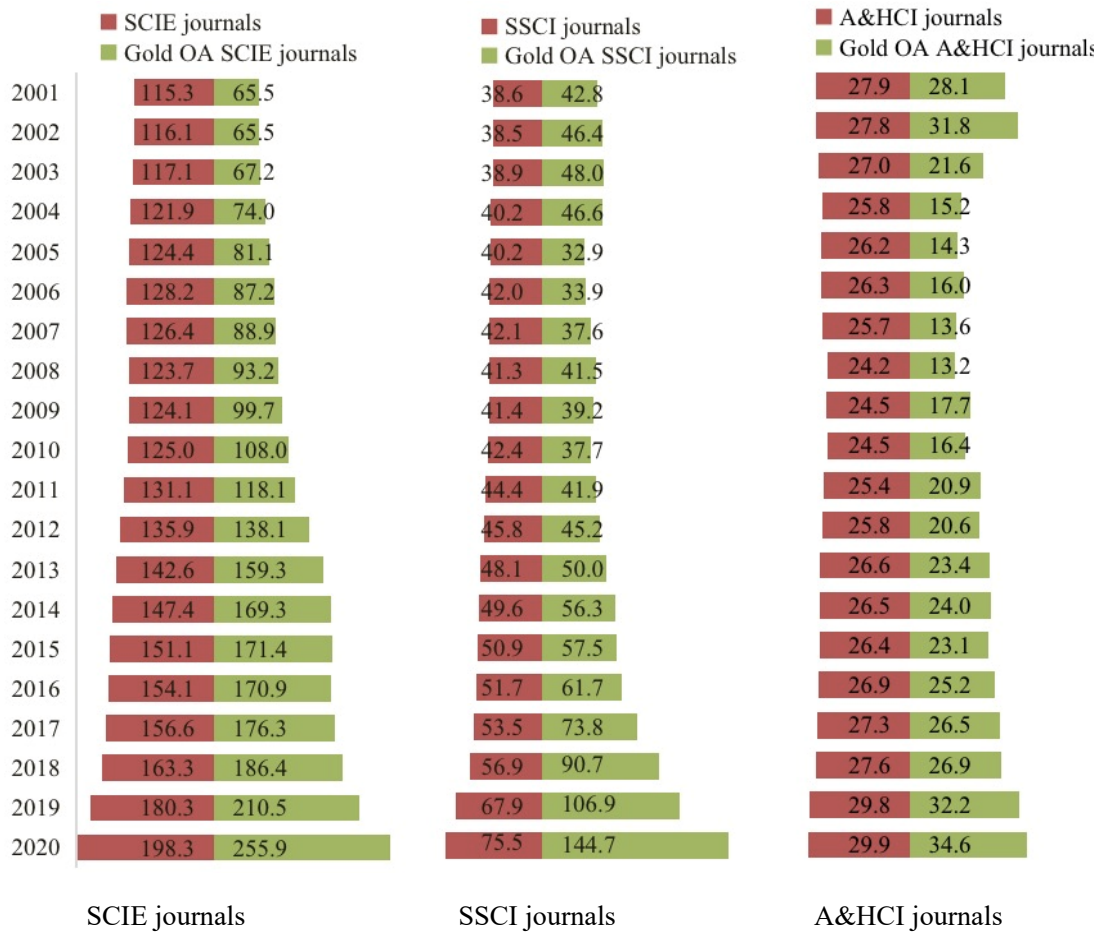

(b) Articles and review articles

**Fig. 9.** Dynamics of average JPVs for Gold OA SCIE/SSCI/A&HCI journals

## 4. Conclusions and discussions

### 4.1 Main findings

Utilizing data from WoSCC's three journal citation indexes, this study uncovered the unbalanced expansion patterns of these three databases from 2001 to 2020. Our findings showed that, in general, the annual numbers of papers in the SCIE and SSCI databases have increased steadily. However, annual publication volumes in the A&HCI database have remained relatively stagnant when considering all document types over the past 20 years. In another situation, if only citable items are considered, not only have the numbers of citable items in the SCIE and SSCI databases increased significantly, but also the number of citable items in the A&HCI database has gained a slight increase over the past 20 years. Similar trends can also be witnessed after excluding individually selected items in the SSCI/A&HCI database.

Echoing the trends of publication volumes, publication titles retrieved from SCIE/SSCI database have also followed upward trends except for the publication titles retrieved from A&HCI database. However, after excluding individually selected journals, all the numbers of SCIE/SSCI/A&HCI journals followed upward trends, especially during the period of 2006-2010. Among them, SSCI journals had the highest growth rate, followed by A&HCI journals and SCIE journals.

Except for the continuous inclusion of new publication titles into citation indexes, changes in journal publication volume are also an important factor contributing to the publication volume expansion. Both the average JPVs of SCIE and SSCI journals have increased gradually, especially during the last three years (for both all document types considered and only citable items considered scenarios). It is partially due to the emergence of large-volume open access SCIE/SSCI journals. Surprisingly, the average JPVs of A&HCI journals decreased gradually for all document types considered scenario or kept relatively stagnant for citable items considered scenario. The small share of open access A&HCI journals and their relatively small publishing volumes does not benefit the increase of the average JPVs of A&HCI journals.

**4.2 Discussions**

It is worth mentioning that both SSCI and A&HCI databases contain significant proportions of papers from individually selected journals. The data in this paper also show that the exclusion of papers from individually selected journals has an important impact on the relevant research conclusions. In some research evaluation scenarios, we need to only focus on the SCIE/SSCI/A&HCI journals and corresponding papers by excluding individually selected ones. The proposed journal reclassification method in this article can identify and exclude individually selected journals/papers quickly and efficiently. Although the latest Journal Citation Reports is also an option to identify SCIE/SSCI/A&HCI journals, a full list of A&HCI journals has not been provided before 2020.

In many bibliometric analysis studies and research evaluation practices, we are familiar with the "growth" for research output of countries/research institutions/research fields and deem it as the "norm". It looks like a win-win-win situation for multiple stakeholders. However, this situation sometimes is just an artifact caused by the expansion of bibliometric databases themselves (Hu, Leydesdorff & Rousseau, 2020; Michels & Schmoch, 2012). Different from the remarkable increase of annual publication volumes in SCIE and SSCI databases, we are surprised to find the near "stagnation" of annual publication volumes in the A&HCI database/all A&HCI journals over the past 20 years (all document types considered). The "stagnation" of annual publication volumes in the A&HCI database/all A&HCI journals is not due to the "stagnation" of the numbers of A&HCI journals. Comparatively, the growth rate of A&HCI journals is only a bit lower than that of SSCI journals but higher than that of SCIE journals. Luckily, the number of articles and reviews in the A&HCI database/all A&HCI journals increase slightly although much more slowly than their SCIE/SSCI counterpart.

This study also selected the sample from the subject area of Arts and Humanities in Scopus as a case for comparative analysis[13]. The number of Arts and Humanities publications in Scopus is nearly constantly growing, with an increase of 392% during the period of 2001-2020. In contrast, as depicted in Fig. 2 (a), the number of publications in the A&HCI database nearly stalled during this period. The number of Arts and Humanities publications in Scopus was only about 30% of that in A&HCI database for the year 2001, however, the number of publications in A&HCI database was only about 70% of that in Scopus for the year 2020. After limiting the country to the USA, the number of Arts and Humanities publications in Scopus has also increased significantly, with an increase of 190% during the past two decades, although corresponding number is almost unchanged in the A&HCI database. Therefore, it would be biased and prejudiced to rely on the A&HCI database to map and evaluate the evolution of the arts and humanities disciplines.

---

[13] Each database has its own coverage. Which database should be more likely to be the gold standard is not the scope of our study. By comparing with Scopus via three cases, it can at least prove that the stagnation of the number of papers in A&HCI is not a normal phenomenon that can be ignored.

Three factors can account for the above "stagnation" or "slight increase" of annual publication volume in the A&HCI database. Although the growth rate of the number of A&HCI journals is a bit higher than that of SCIE journals but much lower than that of SSCI journals, the average annual journal publication volume of A&HCI journals decreases remarkably (all document types considered) or keep relatively stable (only articles and reviews considered). The increase in the number of A&HCI journals may also be partially offset by the increasing paper length. Besides, many mega OA SCIE/SSCI journals benefit the recently remarkably increasing publication volumes of SCIE/SSCI journals, however, the small share of open access A&HCI journals and their relatively small publishing volumes does not benefit the growth of total publication volume of A&HCI journals. Moreover, it is worth mentioning that the sudden shift of individually selected policies in the A&HCI database since 2018 also serves as one factor.

Web of Science Core Collection is a widely used data source for various bibliometric analyses and research evaluation activities (Li et al., 2018). Compared to SCIE/SSCI, the "stagnation" or "slight increase" of annual publication volume in the A&HCI database/all A&HCI journals may influence the above activities (Huang & Chang, 2008; Hicks & Wang, 2014; Franssen & Wouters, 2019). Firstly, the "norm" of publication volume growth found in many natural and social science fields may not be evident or disappear directly in arts and humanities-related disciplines if A&HCI is used as the data source. Compared with natural and social science fields, the contribution from arts and humanities-related disciplines may be underestimated. Secondly, if the number of researchers in arts and humanities-related disciplines also increases, these researchers have increasingly fewer opportunities to publish in A&HCI journals. Since papers indexed by prestigious citation indexes is a widely used proxy indicator of high-quality work used in researcher promotion, funding grant, and institution ranking in some countries, arts and humanities-related researcher/research fields/institutions will be hindered. Therefore, the unbalanced development of three citation indexes in WoSCC will weaken the diversified development of disciplines and pose a more significant threat to the progress of arts and humanities disciplines, especially in countries that value quantitative indicators (Vanderstraeten & Vandermoere, 2021).

Inequality rather than scarcity is the root of trouble. The "stagnation" or "slight increase" of annual publication volume in the A&HCI database/all A&HCI journals indicates that efforts still should be taken by many stakeholders. For the data provider of the Web of Science Core Collection, Clarivate should continue to index more high-quality journals and books from arts and humanities disciplines. For journal publishers of arts and humanities disciplines, they should try to break the page limit of each issue, including more aggressively embracing the online open access publishing model. For various evaluators and management personnel, the facts about the A&HCI database/all A&HCI journals should be clearly understood and used wisely. Some other additional databases are recommended for research evaluation in arts & humanities disciplines. As scientometricians, we should also take the initiative to explore the secrets behind many prestigious data sources and disseminate them to stakeholders as widely as possible.

**4.3 Limitations and future work**

A research limitation of this study is the lack of a gold standard to confirm whether the citation index is stagnant or not. Scopus also has its coverage limitations. It may be compared to other databases (such as institutional repositories) in the future to find a gold standard for comparison. Another caveat of this study is that the analysis dimension is relatively macro, and the expansion mechanisms of different sub-disciplines under the A&HCI database may vary substantially. Moreover, this study did not consider other key types of output (e.g. books) in the fields of arts & humanities, which would be worthy of further investigation in the future.